\documentclass[twocolumn,showpacs,showkeys]{revtex4}
\usepackage{epsfig}
\usepackage{graphicx}
\usepackage{bm}
\usepackage{amsmath}
\usepackage{amsfonts}

\newtheorem{claim}{Claim}
\newtheorem{thrm}{Theorem}

\begin{document}
\title{Playing The Hypothesis Testing Minority Game In The Maximal Reduced
 Strategy Space}
\author{H.~F. Chau, V.~H. Chan and F.~K. Chow} 
\affiliation{ Department of Physics and Center of Theoretical and Computational
Physics, \\
University of Hong Kong, Pokfulam Road, Hong Kong}

\begin{abstract}
Hypothesis Testing Minority Game (HMG) is a variant of the standard Minority
Game (MG) that models the inertial behavior of agents in the market.  In the
earlier study of our group, we find that agents cooperate better in HMG than in
the standard MG when strategies are picked from the full strategy space. Here
we continue to study the behavior of HMG when strategies are chosen from the
maximal reduced strategy space. Surprisingly, we find that, unlike the standard
MG, the level of cooperation in HMG depends strongly on the strategy space
used. In addition, a novel intermittency dynamics is also observed in the
minority choice time series in a certain parameter range in which the orderly
phases are characterized by a variety of periodic dynamics. Remarkably, all
these findings can be explained by the crowd-anticrowd theory.
\end{abstract}


\maketitle

\section{introduction}
Studying economic systems by agents-based models have attracted the attention
among physicists in recent years \cite{AAP88,M89,Z98}. One of the most famous
agents-based model in this regard is the Minority Game (MG) \cite{Z98,CZ97}.
MG does not only capture the fact that all people in the market think
inductively and selfishly \cite{A94}, its complexity also satisfy the definition
of a complex system in the strictest sense \cite{C05}. In spite of its simple
governing rules, agents in this model self-organize giving rise to an
unexpected global cooperative phenomena.

Using the standard MG as blue print, various {modifications} to the rules of the
standard MG have been proposed \cite{GMS2000,CC03,CGGS99,LL06} to understand
different aspects and phenomena in realistic economic systems. In particular,
Man and Chau introduced the Hypothesis Testing Minority Game (HMG) to model the
inertial behavior of agents \cite{MC06}. They found that the presence of
inertial agents improve global cooperation leading to a decrease of the variance
per agent over the entire parameter space provided that the strategies of each
agent are chosen from the so-called full strategy space (FSS) \cite{MC06,CM07}.

In this paper, we move on to study the agent cooperation and the dynamics of
HMG in case the strategies are picked from the so-called maximal reduced
strategy space (MRSS) \cite{CZ98}. We begin by briefly reviewing the rules of
HMG and stating the parameters used in our numerical simulations in
Sec.~\ref{Sec:Game}. Then we report our simulation results in
Sec.~\ref{Sec:Result}. To our surprise, the behavior of HMG depends strongly on
the strategy spaces used. Specifically, agents generally cooperate better when
strategies are picked from the FSS rather than the MRSS provided that they are
sufficiently reluctant to change their strategies.
In contrast, the standard MG is so robust that its dynamics and cooperative
behavior are essentially the same irrespective of whether the FSS or the MRSS
is used. Furthermore, we find that in HMG the minority
choice time series exhibits intermittency in which the orderly phases show
periodic dynamics with period up to $2(2^M-1)$ whenever the memory size of the
strategies $M$ is greater than $1$ in a certain parameter range
when strategies are picked from the MRSS. This novel intermittent phenomenon
does not show up in HMG provided that strategies are
picked from the FSS as well as in the standard MG. We explain how these
differences originate from the
choice of the strategy space by a semi-analytical approach known as the
crowd-anticrowd theory \cite{HJHJ01,HJHJ01a} in Sec.~\ref{Explain}.
In fact, the major reason responsible for these differences is
that it is a lot easier for an agent to
keep on using one's currently adopted strategy when the strategy pool is
the MRSS than rather than FSS in certain parameter regime.
Finally, we summarize our findings in Sec.~\ref{Final}.
Our findings show that extra care is
needed to study variants of MG as their behavior may depend sensitively on the
strategy space employed.
Nonetheless, the ability to explain the behavior of HMG using the
crowd-anticrowd theory suggests that this theory may still be useful to
explain the dynamics of variants of the standard MG provided that one
carries out the analysis carefully.

\section{Hypothesis Testing Minority Game}\label{Sec:Game}
Recall that in the standard MG, agents act according to the predictions of their
best performing strategies. In other words, agents in the standard MG do not
hesitate to stop
using their current strategies once the performance indicator, known as virtual
score, shows that the strategies are not the  best. In contrast, the HMG
incorporated the inertial behavior of agents by allowing them to stick to their
currently using strategies until their performances are too poor to
be acceptable.
More precisely, a fixed real number $I_k$ between 0.5 and 1.0 is assigned once
and for all to each agent~$k$ in HMG to represent their reluctance to switch
strategies. Using the value of $I_k$ as an indicator of the confidence level,
agent~$k$ tests the hypothesis that his currently using strategy is his best
strategy at hand at each turn. Furthermore, he switches to another
strategy and resets the virtual scores of all his strategies to $0$ if
the null hypothesis is rejected. Apart from these differences, the
governing rules of HMG are identical to those of the standard MG. 

We state the rules of HMG below for reader's convenience. 

\subsection{Rules of the game}

\begin{enumerate}
\item HMG is a repeated game of a fixed population of $N$ agents. 
A number $I_k \in [0.5,1) $ is assigned to agent~$k$ once and for all to
represent his inertia.
\item At each turn $\tau$, every agent has to make a choice between one of 
the two sides (namely side $0$ and side $1$) based on the strategies to be
described in rule~\ref{HMGrule4}.
Those agents in the side with the least number of agents (known as the minority
side) win in that turn. 
And in case of a tie, the winning side is randomly selected.
\item The only piece of global information reveals to the agents at time $\tau$ is the 
winning sides in the last $M$ turns known as the history $\vec{\mu}(\tau)$.
\item Before the game commences, each agent~$k$ is assigned once and for all
$S$ randomly picked strategies $S_{k,i}$ for $i = 0,1,\ldots ,S-1$. Each
strategy $S_{k,i}$ is a function map 
from the set of all possible histories to the set $\{0,1\}$ and its virtual
score $\Omega_{k,i}$ is set to $0$ initially.
Without loss of generality, strategy $S_{k,0}$ is assumed to be the currently
using strategy of agent~$k$ at the beginning of game.
\label{HMGrule4}
\item Agent~$k$ will switch his current strategy from $S_{k,0}$ to $S_{k,j}$ if
and only if the maximum virtual score difference $\Delta \Omega_k $ drops below
the threshold $x_k \sqrt{2\tau_k}$, that is, 
\begin{eqnarray}
\Delta \Omega_k & \equiv & \max_i  \{\Omega_{k,0} - \Omega_{k,i}\} =
\Omega_{k,0} - \Omega_{k,j} \nonumber \\
& \leq & x_k \sqrt{2\tau_k} \label{EDOmega}
\end{eqnarray}
where $x_k$ is defined by 
\begin{equation}
\frac{1}{\sqrt{2\pi}} \int^{+\infty}_{x_k} e^{-x^2/2} \,dx = I_k 
\end{equation}
and $\tau_k$ is the number of turns elapsed since agent~$k$'s last
switch of strategy.
In case agent~$k$ switches his strategy, he exchanges the labels $0$ and $j$
so that his currently using strategy is always labeled as $S_{k,0}$.
In addition, the virtual scores $\Omega_{k,i}$ are reset to 0 for all $i$ and
$\tau_k$ is reset to 1.
\label{HMGrule5}

\item Agent~$k$ uses his current strategy to guess the minority choice of
the current turn. Moreover, the virtual score of strategy $S_{k,i}$ is
increased (decreased) by $1$ if it predicts the minority side of that turn
correctly (incorrectly).
\end{enumerate}

\subsection{Parameters used in our simulation}
\label{Subsec:Parameters}

We select the following parameters in our simulations:
\begin{enumerate}
\item $N$ is odd;
\item $S=2$;
\item all values of $I_k$ are chosen to be the same independent of the agent
label $k$ (and we write this common $I_k$ as $I$ for simplicity); and
\item $S_{k,j}$ are picked from the so-called MRSS \cite{CZ98}. (That
is, $S_{k,j}$ can be written in the form
\begin{equation}
 s_\tau = \eta_0 + \sum_{i=1}^M \eta_i \mu_{\tau-i} ~, \label{E:MRSS_Def}
\end{equation}
where $s_\tau$ is the prediction of the minority side in the $\tau$th turn,
$\eta_0,\eta_1,\ldots , \eta_M \in \{ 0,1 \}$, $\mu_i$ is the minority side in
the $i$th turn and the arithmetic is performed in the finite field of two
elements $GF(2)$ \cite{CZ98,CC03,CC04}. In other words, a strategy in the MRSS
is characterized by $(\eta_0,\eta_1, \ldots ,\eta_M)$.)
\label{Parameter4}
\end{enumerate}
With the exception of
point~\ref{Parameter4}, the parameters used in this study are
identical to those used in our earlier study of HMG reported in
Refs.~\cite{MC06,CM07}. In contrast, strategies in Refs.~\cite{MC06,CM07} are
picked from the FSS, namely, the set of all possible strategies. And a strategy
in the FSS may not be expressed as a linear function of $\mu_i$'s. From now on,
we use the symbols HMG$^\text{FSS}$ and HMG$^\text{MRSS}$ to denote the HMG in
which strategies are picked from the FSS and the MRSS, respectively.

\section{Our Numerical Simulation Results}\label{Sec:Result}

\subsection{Focus of our study}

We are interested in both the cooperative behaviors and the dynamics of the
game. Recall that
MG and HMG are non-positive sum games in the sense that the number of winning
agents is less than or equal to the number of losing agents in each turn.
And we say that the agents (or the system) cooperate better if the average
number of winning agents per turn is high. Our numerical simulations
show that agents self-organize in such a way that there is no bias in picking
the minority side when averaged over the agents and the number of turns,
so we follow the usual practice to
study agent cooperation by means of the $\alpha\equiv 2^{M+1} / NS$ against
$\sigma^2 / N$ graph where $\sigma^2$ is the variance of the number of agents
choosing side~0 \cite{CZ98}. The lower
the value of $\sigma^2 / N$, the better the agent cooperation.

As for the dynamics of HMG, our investigation focuses on the analysis of
the periodicity of the
minority choice time series through the auto-correlation function.
And the auto-correlation function can be conveniently studied by means of a
time lag $t$ against auto-correlation $C_0$ graph. In order to
make sure that the dynamics is genuine and long lasting, we only consider the
time series after the system has equilibrated. We also perform simulations
using different values of $N$ and initial quenched disorders to make sure that
the dynamics we are going to report below are generic. 

Actually, the dynamics depends on the following three factors:
\begin{enumerate}
\item number of agents $N$;
\item history size $M$; and
\item the initial quenched disorder as reflected by the value $I$ and the
strategies $S_{k,i}$ assigned to the agents. 
\end{enumerate}
Our choice of parameters for the HMG$^\text{MRSS}$ reported in
Sec.~\ref{Subsec:Parameters} makes the dynamics
of the game deterministic and hence enabling us to study the periodic
dynamics of the minority choice time series easily.
In contrast, when played using other
choices of $N$ and $S$, the non-deterministic nature of this game weakens the
periodic dynamics in minority choice time series, making both the numerical and
analytical studies more troublesome.  

Unlike the standard MG, we find that both the cooperative behavior and the
dynamics of HMG depend strongly on the strategy space chosen. We shall
elaborate more on this point in the coming two subsections. 

\begin{figure}[ht]
\centering\includegraphics[width=7.0cm]{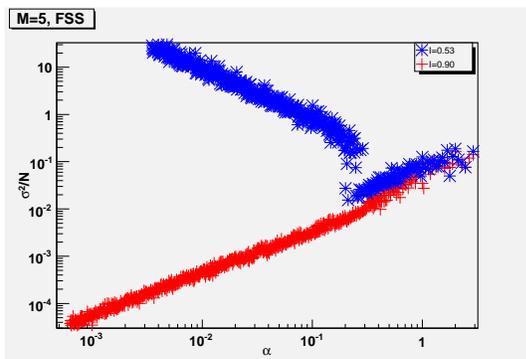}
\caption{The $\alpha$ against $\sigma^2 / N$ graph for HMG$^\text{FSS}$. The
 value of $\alpha$ is varied by fixing $M$ and changing $N$.}
\label{FCoopFSS}
\end{figure}

\begin{figure}[ht]
\centering\includegraphics[width=7.0cm]{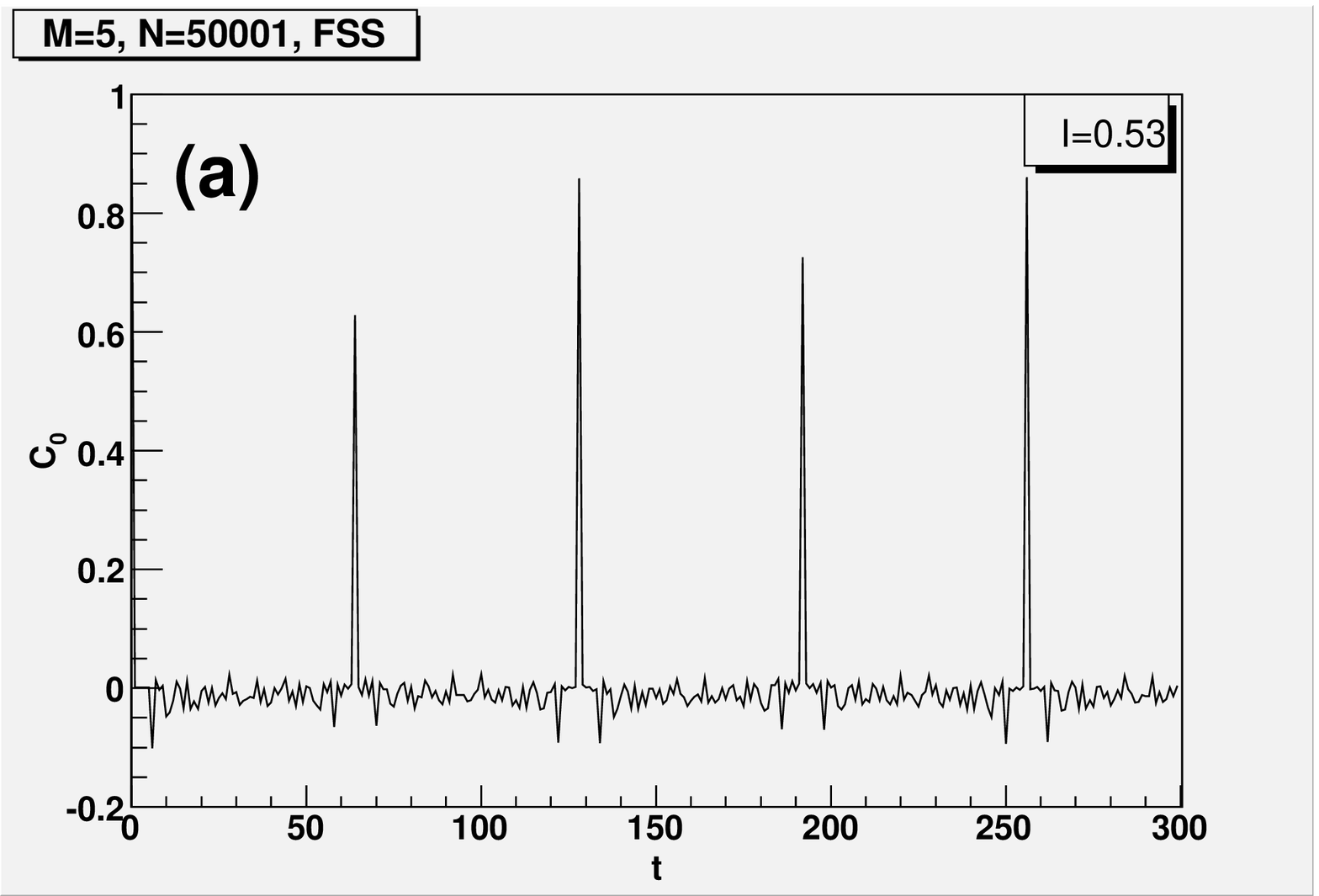}
\centering\includegraphics[width=7.0cm]{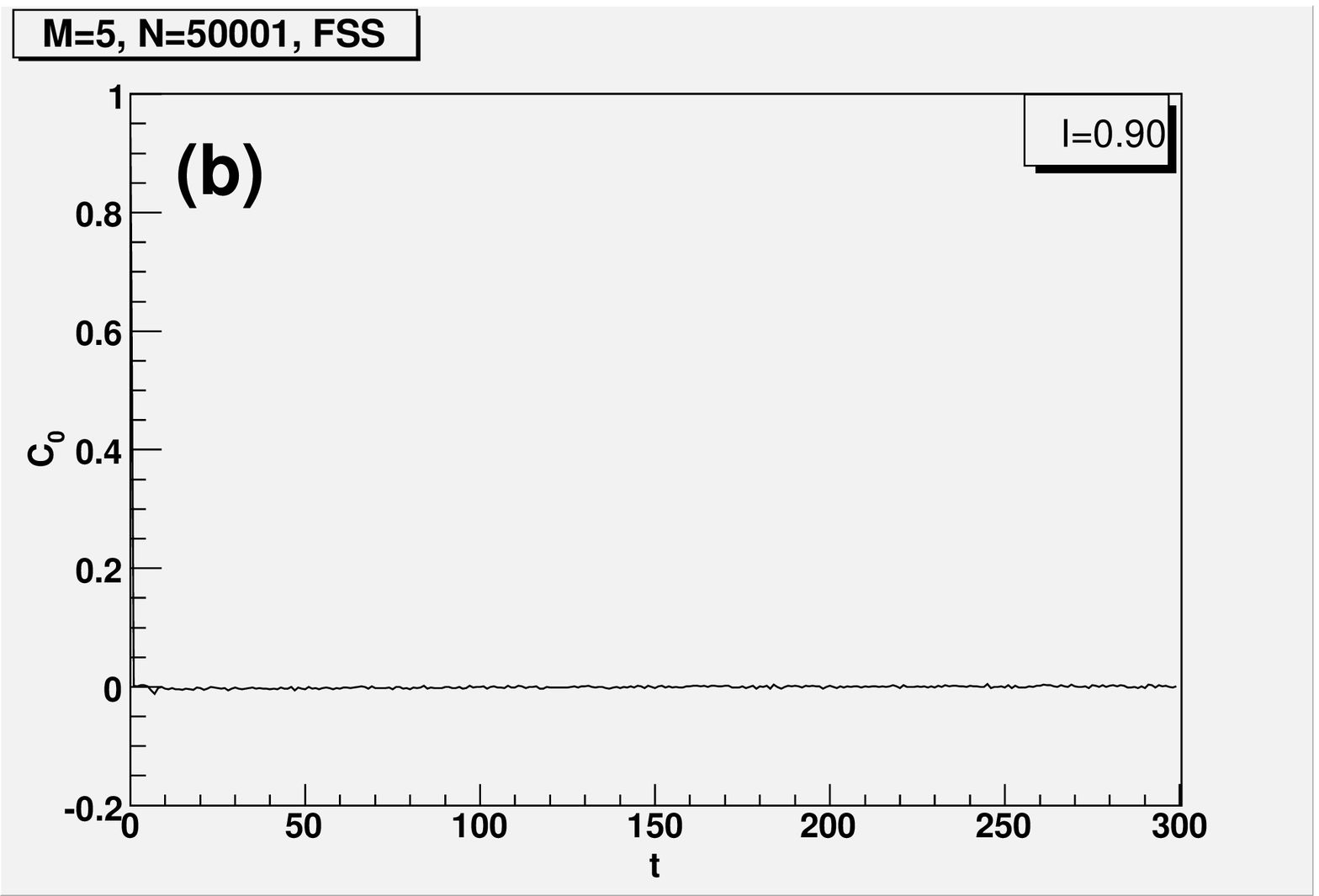}
\caption{The auto-correlation for HMG$^\text{FSS}$ with $\alpha =
 6.4\times 10^{-4} \ll \alpha^\text{MG}_c$ when (a)~$I = 0.53$ and
 (b)~$I = 0.90$.}
\label{FDynFSS}
\end{figure}

\subsection{Reviewing the simulation results of the HMG$^\text{FSS}$}
Recall from the earlier study of our group in Refs.~\cite{MC06,CM07} that when
$I$ is chosen to be less than $I^{c_1}$, where
\begin{equation}
 I^{c_1} = \frac{1}{\sqrt{2\pi}} \int^{+\infty}_{-\sqrt{2/2^{M+1}}} e^{-x^2/2}
 \,dx ~, \label{I_c_def}
\end{equation}
the inertia of agents is not strong enough to make the dynamics of the
HMG$^\text{FSS}$ to deviate significantly from that of the standard
MG. Thus, the $\sigma^2 / N$ is about the same as that of the standard MG.
Moreover, the well-known period $2^{M+1}$ dynamics in the minority
choice time series that appears in the standard MG when $\alpha$ is
less than $\alpha^\text{MG}_c \approx 0.3$, the critical value for the standard
MG, is also present here \cite{MC06,CM07}. In contrast, when
$I > I^{c_1}$, the inertia
of agents becomes strong enough to significantly reduce the herd effect amongst
agents resulting in a much lower $\sigma^2 / N$ (and hence indicating that
agents cooperate better). Besides, the period $2^{M+1}$ dynamics is no longer
present when $\alpha < \alpha^\text{MG}_c$ \cite{MC06,CM07}. These earlier
findings are summarized in Figs.~\ref{FCoopFSS} and~\ref{FDynFSS}.

\begin{figure}[hb]
\centering\includegraphics[width=7.0cm]{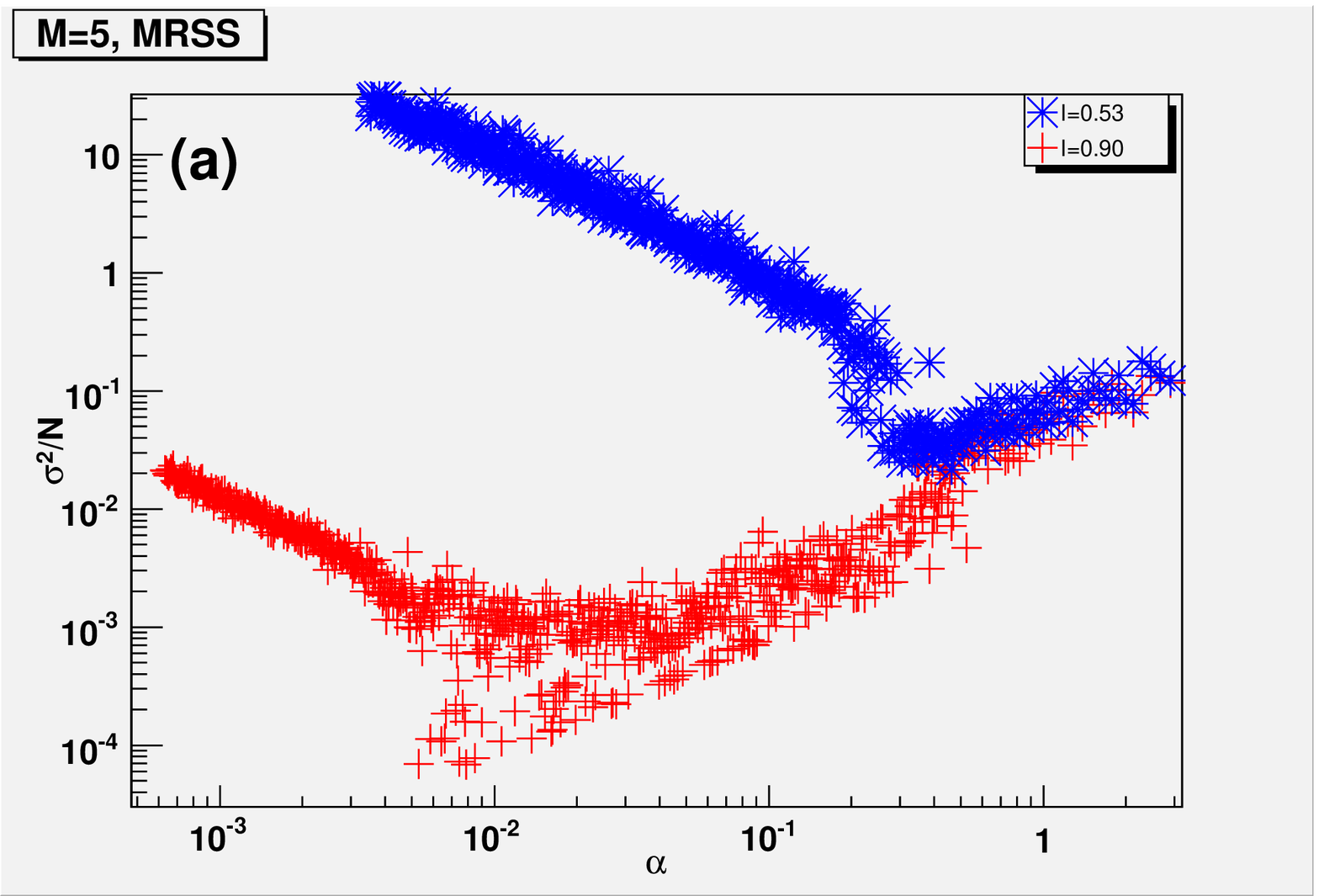}
\centering\includegraphics[width=7.0cm]{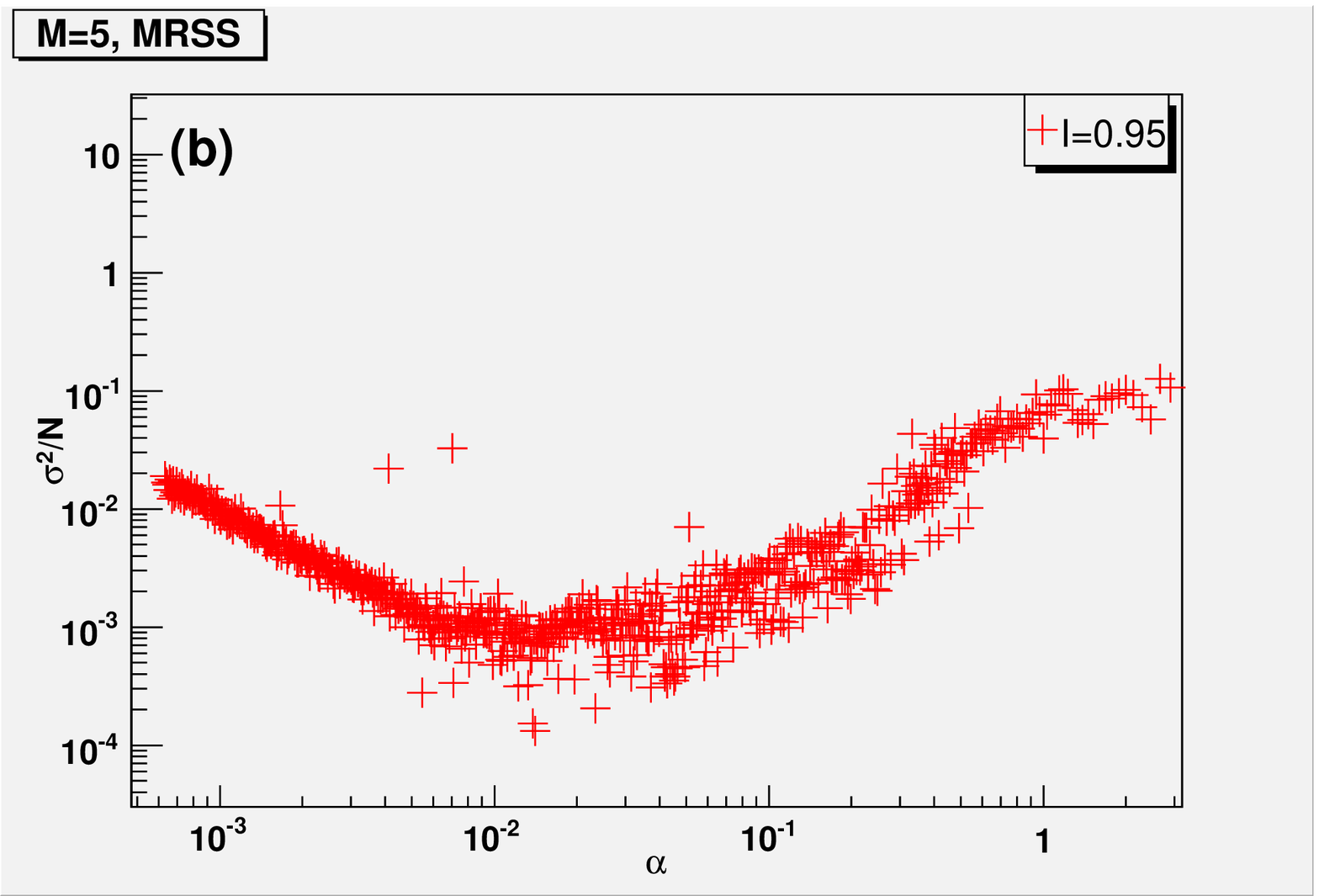}
\caption{Plots of $\alpha$ against $\sigma^2 / N$ for HMG$^\text{MRSS}$ with
 (a)~$I = 0.53, 0.90$ and (b)~$I = 0.95$.}
\label{FCoopMRSS}
\end{figure}

\subsection{Simulation results of the HMG$^\text{MRSS}$}
Contrary to our expectation, we find that the behavior of
HMG$^\text{MRSS}$ is significantly different from that of HMG$^\text{FSS}$ when
$I \gtrsim I^{c_1}$ and $\alpha \lesssim 1$. The details of our findings are
listed below.

\begin{figure}[ht]
\centering\includegraphics[width=7.0cm]{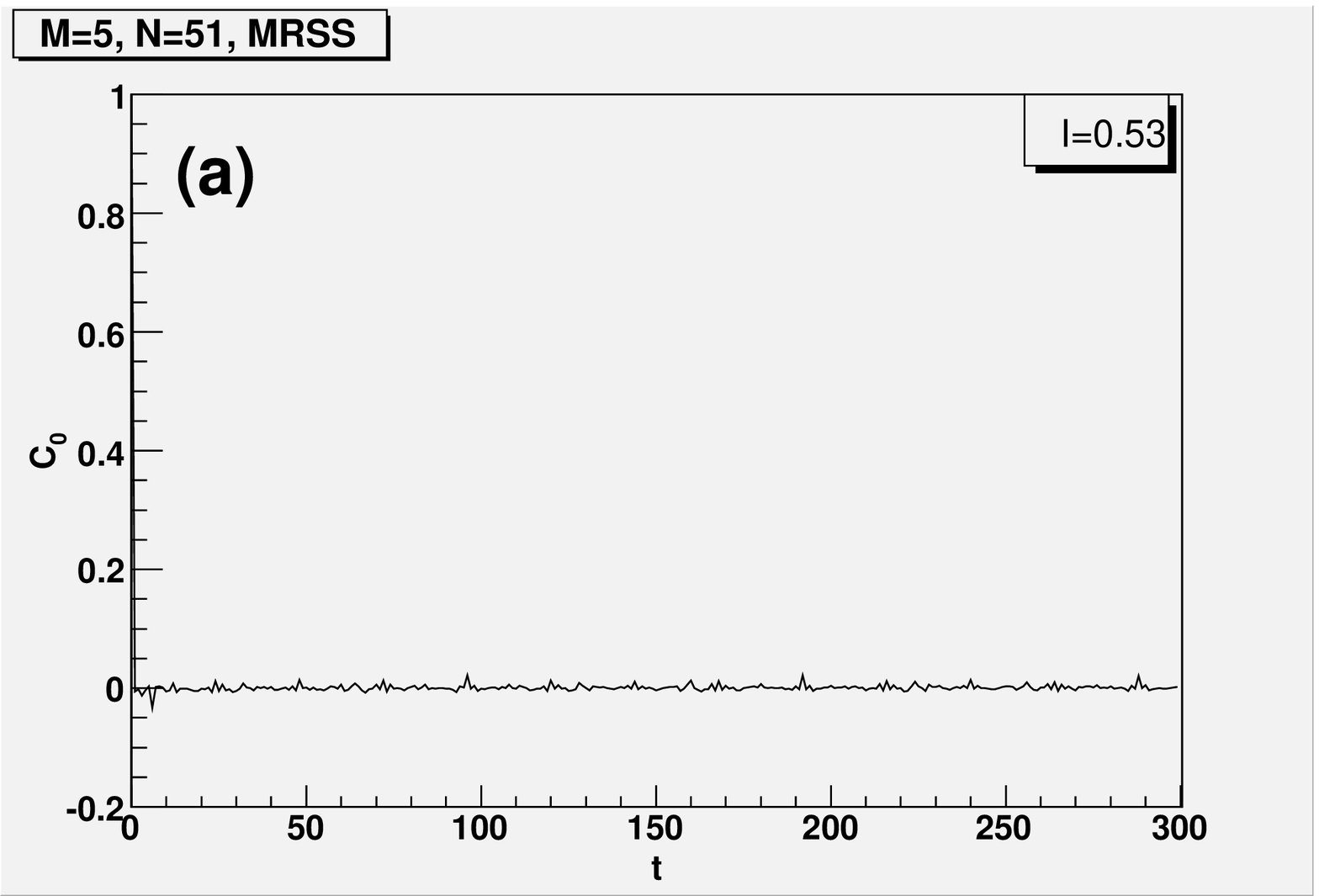}
\centering\includegraphics[width=7.0cm]{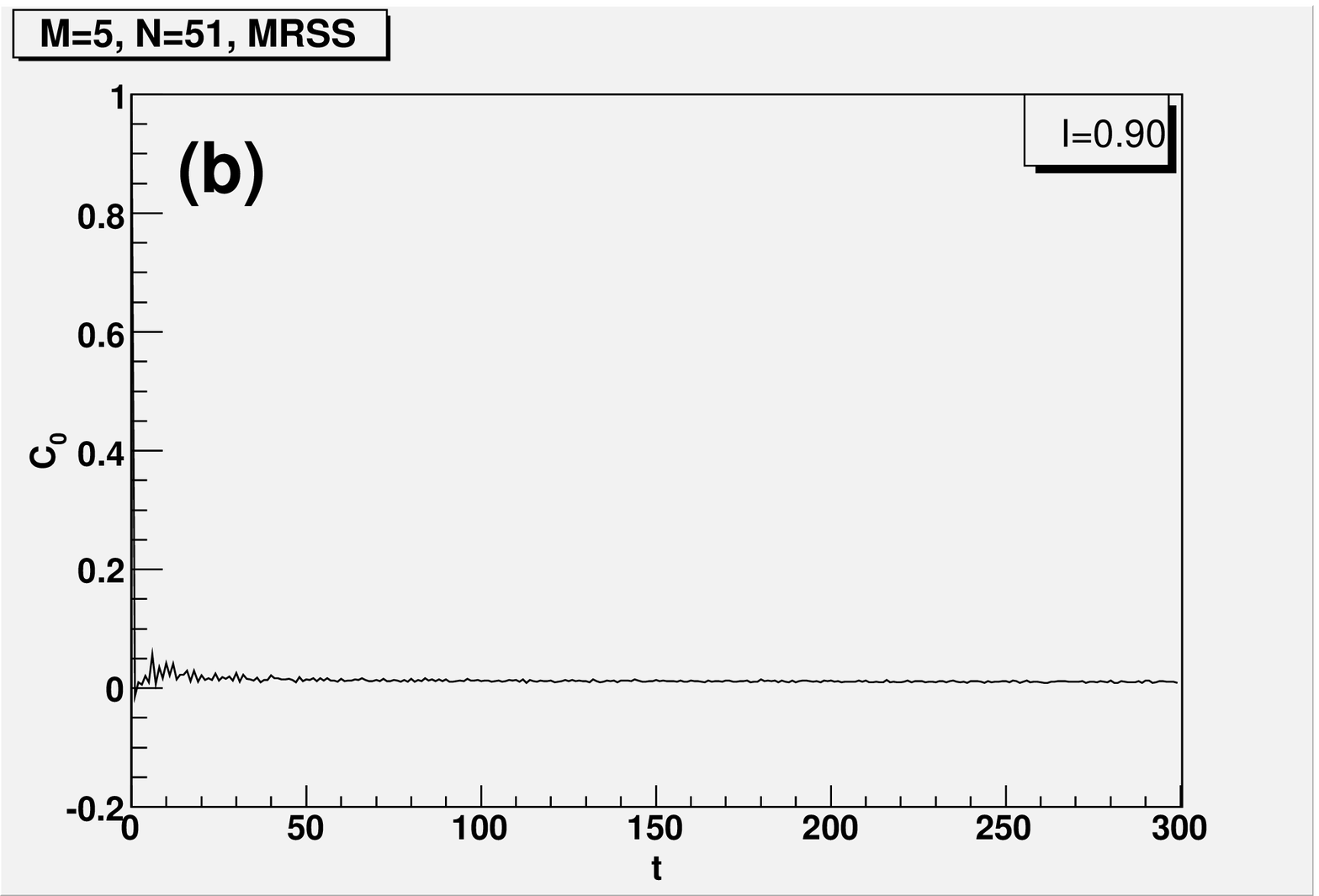}
\caption{The auto-correlation of HMG$^\text{MRSS}$ for
 $\alpha = 0.63 \gg \alpha^\text{MG}_c$ when (a)~$I = 0.53$ and (b)~$I =
 0.90$.}
\label{FDynMRSS_Ba}
\end{figure}

\begin{figure}[ht]
\centering\includegraphics[width=7.0cm]{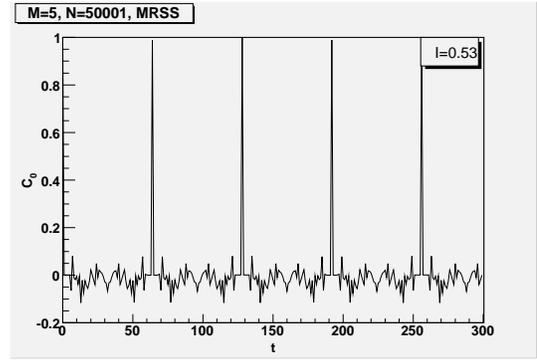}
\caption{The auto-correlation of HMG$^\text{MRSS}$ for $I < I^{c_1}$ and
 $\alpha = 6.4\times 10^{-4} \ll \alpha^\text{MG}_c$.}
\label{FDynMRSS_SISa}
\end{figure}

\begin{figure}[ht]
\centering\includegraphics[width=7.0cm]{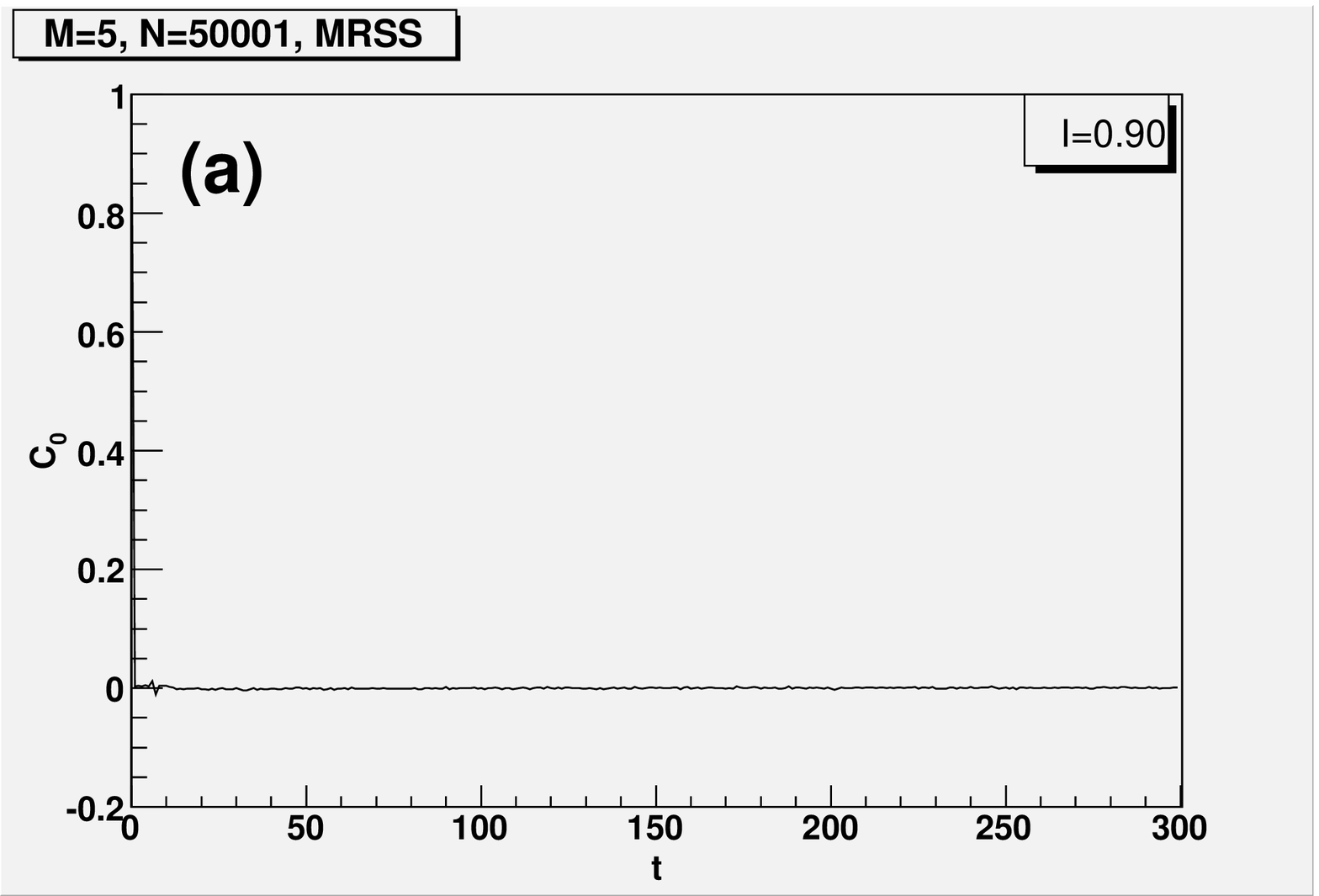}
\centering\includegraphics[width=7.0cm]{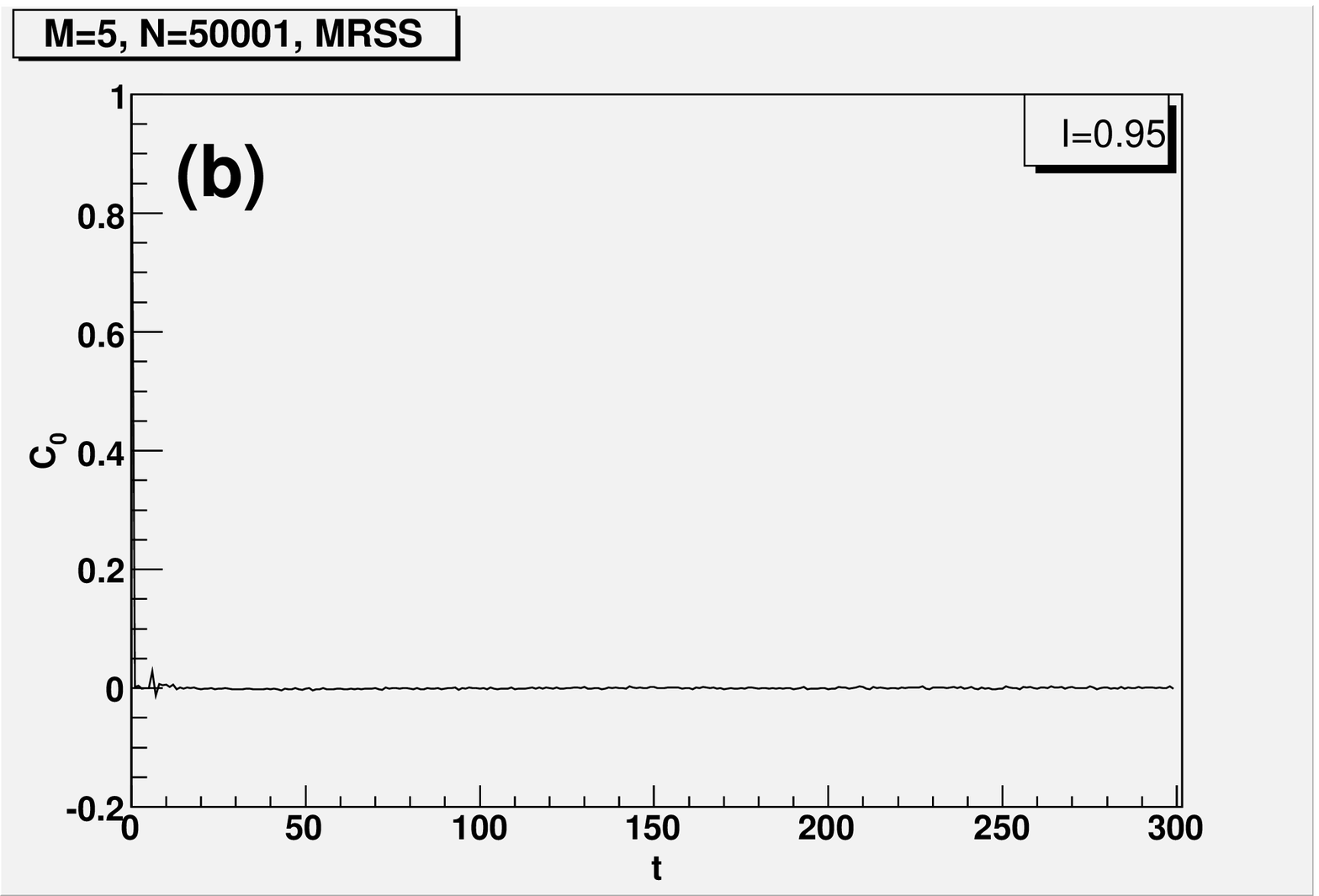}
\caption{The auto-correlation of HMG$^\text{MRSS}$ when $\alpha = 6.4\times
 10^{-4} \ll \alpha^\text{MG}_c$ together with (a)~$I^{c_1} < I < I^{c_2}$ and
 (b)~$I \geq I^{c_2}$.}
\label{FDynMRSS_Sa}
\end{figure}

\begin{figure}[ht]
\centering\includegraphics[width=7.0cm]{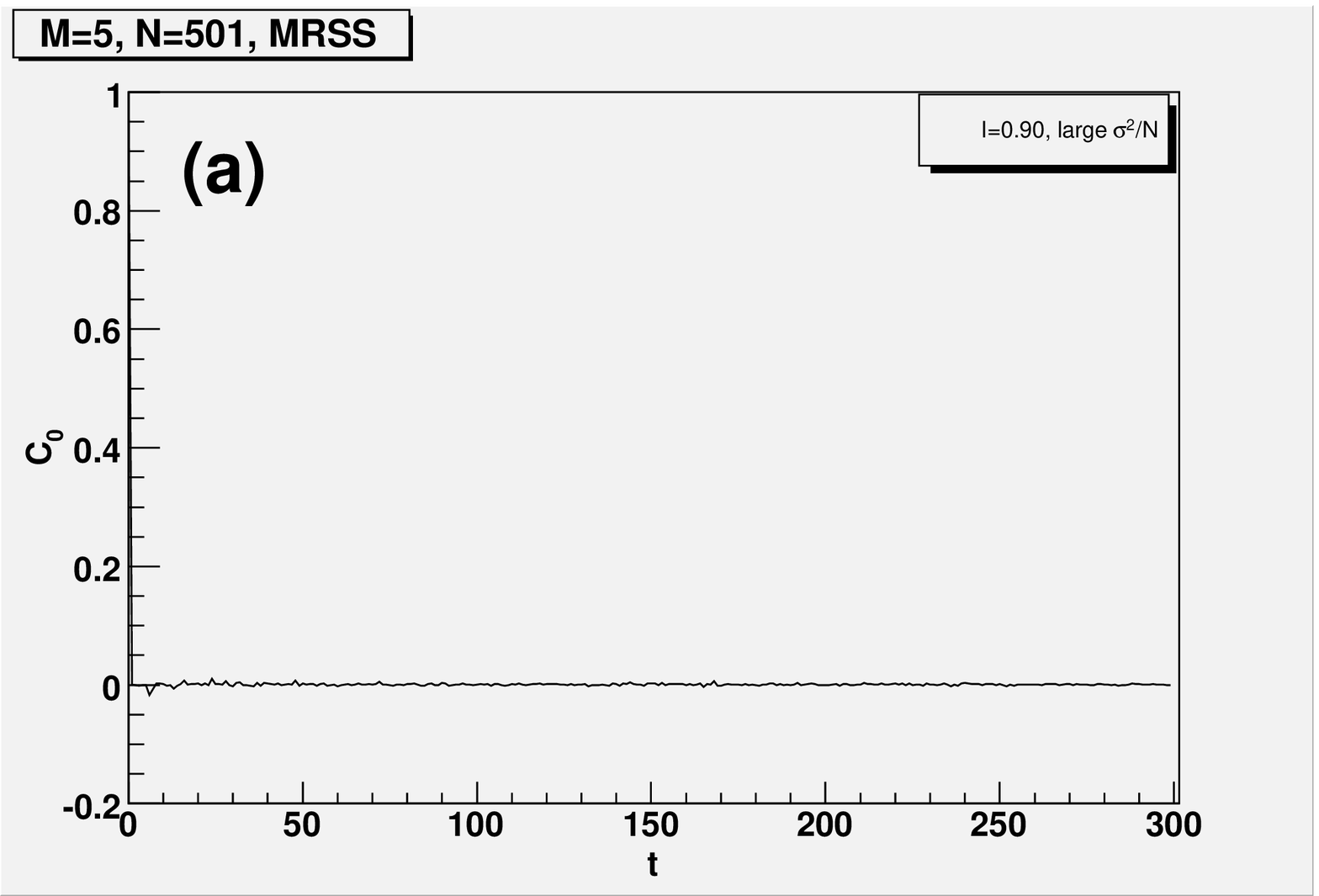}
\centering\includegraphics[width=7.0cm]{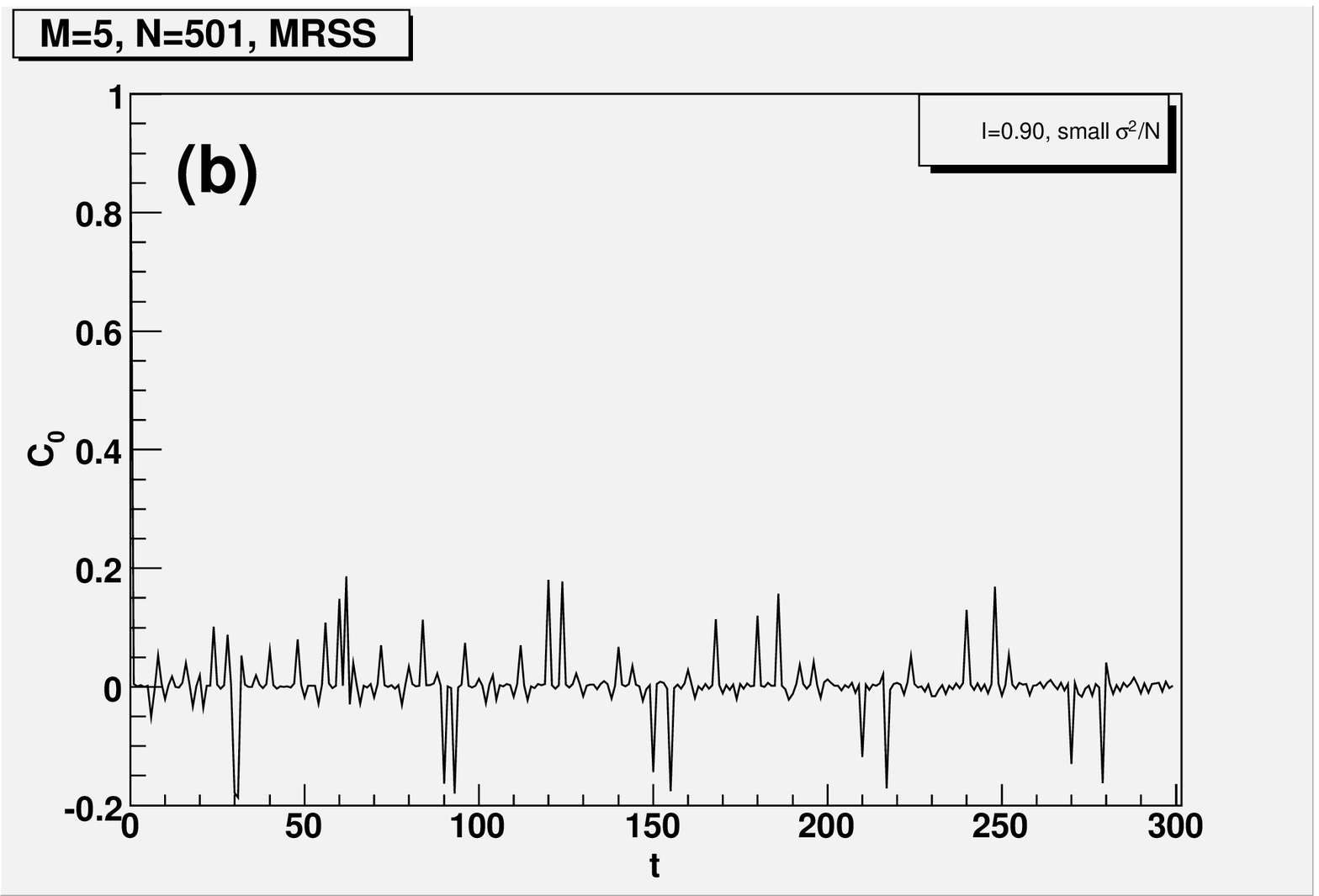}
\centering\includegraphics[width=7.0cm]{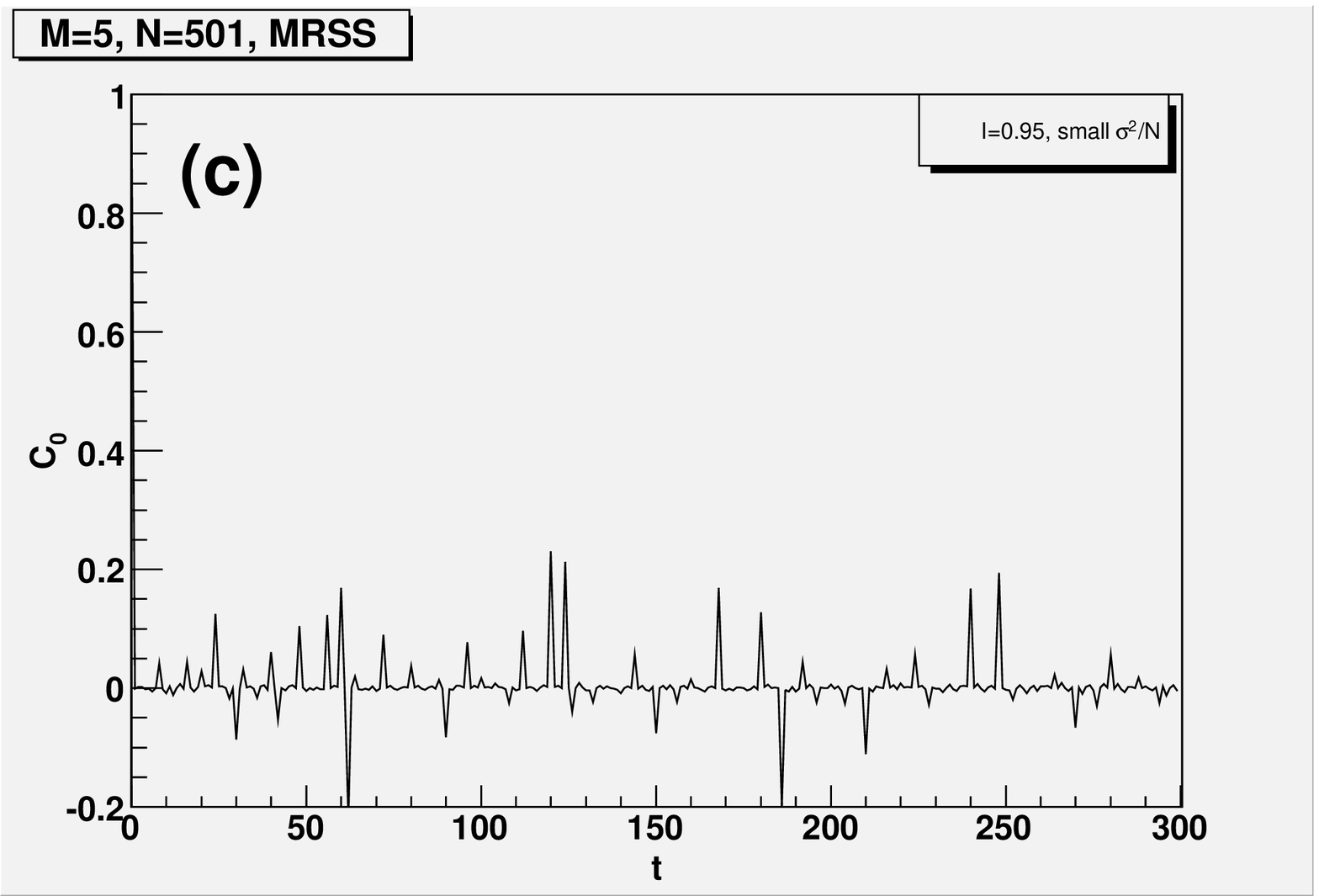}
\caption{The auto-correlation of HMG$^\text{MRSS}$ for $\alpha = 0.064 \approx
 \alpha^\text{MG}_c$ with
 (a)~$I^{c_1} < I < I^{c_2}$ and $\sigma^2 / N$ is large,
 (b)~$I^{c_1} < I < I^{c_2}$ and $\sigma^2 / N$ is small, and
 (c)~$I \geq I^{c_2}$ and $\sigma^2 / N$ is small.}
\label{FDynMRSS_Ma}
\end{figure}

\begin{itemize}
\item $I \lesssim I^{c_1}$: By comparing Fig.~\ref{FCoopFSS} with
 Fig.~\ref{FCoopMRSS}a and Fig.~\ref{FDynFSS} with Figs.~\ref{FDynMRSS_Ba}
 and~\ref{FDynMRSS_SISa}, we know that the behavior of HMG$^\text{MRSS}$ in
 this regime is similar to that of HMG$^\text{FSS}$. That is, they have about
 the same level of agent cooperation. In addition, the standard MG
 \cite{HMCC05,CCHM05,MHCC06}, HMG$^\text{FSS}$ \cite{MC06} and
 HMG$^\text{MRSS}$ all exhibit the same period $2^{M+1}$ dynamics in the
 minority choice time series whenever $\alpha \lesssim \alpha^\text{MG}_c$ and
 show no periodic dynamics for $\alpha \gtrsim \alpha^\text{MG}_c$.

\item $I \gtrsim I^{c_1}$ and $\alpha \geq 1 (\gg \alpha^\text{MG}_c)$: By
 comparing Fig.~\ref{FCoopFSS} with Fig.~\ref{FCoopMRSS}, we find that the
 values of $\sigma^2 / N$ are about the same for both HMG$^\text{FSS}$ and
 HMG$^\text{MRSS}$ in this regime. Moreover, the standard MG \cite{CZ98},
 HMG$^\text{FSS}$ \cite{MC06} and HMG$^\text{MRSS}$ all show no periodic
 dynamics.

\item $I \gtrsim I^{c_1}$ and $\alpha \lesssim 0.1 (\ll \alpha^\text{MG}_c)$:
 Figs.~\ref{FCoopFSS} and~\ref{FCoopMRSS} show that for the same value of
 $\alpha$ in this regime, the cooperation amongst agents for the standard MG is
 the worst, for HMG$^\text{MRSS}$ is in the middle and for HMG$^\text{FSS}$ is
 the best. One interesting feature for HMG$^\text{MRSS}$ is that, unlike
 HMG$^\text{FSS}$, the value of $\sigma^2 / N$ increases as $\alpha$ decreases
 in this regime indicating that agents cooperate less and less as the number of
 agents $N$ increases (and with $M$ and $S$ held fixed). As for the dynamics,
 Fig.~\ref{FDynMRSS_Sa} depicts that the system exhibits no obvious periodic
 dynamics. Nonetheless, its minority choice time series conditioned on an
 arbitrary but fixed history exhibits a very weak period two dynamics.

\item $I \gtrsim I^{c_1}$ and $0.1 \lesssim \alpha \approx \alpha^\text{MG}_c <
 1$: In this regime, we find that the value of $\sigma^2 / N$ obtained after
 equilibration depends on the initial quenched disorder of the system
 indicating the presence of a phase transition point.
 (See Fig.~\ref{FCoopMRSS}.) Actually, the values of
 $\sigma^2 / N$ obtained in many runs are rather close to the theoretical
 minimum of $1/4N$ (which is attained when there are exactly $(N-1)/2$ winning
 agents in each turn) implying that agents cooperate almost perfectly.

 The dynamics of the minority choice time series in this regime is rather
 complex. Actually, no obvious periodic dynamics is observed for those initial
 quenched disorder that ends up with values of $\sigma^2 / N$ about the same as
 those for $I < I^{c_1}$. (See Fig.~\ref{FDynMRSS_Ma}a.) In contrast, those
 ending up with a much smaller $\sigma^2 / N$ show intermittency. (See
 Fig.~\ref{FDynMRSS_Ma}b.) That is to say, when $\sigma^2 / N$ is small, the
 time series exhibits periodic dynamics for some time and then the periodicity
 either suddenly disappears or the period of the dynamics changes. Also the
 brief episode of aperiodicity terminates with the commencement of a new
 periodic dynamics (with possibly a new period).

 Interestingly, the period of the orderly phase for this intermittency depends
 on the value of $I$. In case $I^{c_1} \lesssim I < I^{c_2}$, where
\begin{equation}
 I^{c_2} \equiv \frac{1}{\sqrt{2\pi}} \int_{-\sqrt{2}}^{+\infty} e^{-x^2/2}
 \,dx \approx 0.92135 ~, \label{Ic2_def}
\end{equation}
 the periods are less than or equal to $2(2^M - 1)$ whenever $M\geq 2$.
 More importantly, these periods are in the form $2^j \text{L.C.M.}
 (p_1,p_2,\ldots ,p_M)$ where $j\geq 1$ is an integer, $\text{L.C.M.}$ denotes
 the least common multiple of the $M$ arguments and $p_1, \ldots , p_M$ are
 positive integers dividing $(2^M - 1)$. Clearly, this phenomenon is novel and
 is never found in both the standard MG and HMG$^\text{FSS}$. We observe the
 trend that long period dynamics tends to be more stable in the sense that it
 lasts longer. In fact, the longest periodic dynamics, namely the one with
 period $2(2^M - 1)$, appears to be the most stable. However, being the most
 stable dynamics does not necessarily mean that it must show up for every
 initial quenched disorder. Actually, the period $2(2^M - 1)$ dynamics is
 harder and harder to find as $M$ increases beyond about $7$.

 In the case of $I \geq I^{c_2}$, the periodic dynamics of the orderly phase of
 the intermittency is weak compared with the case of $I^{c_1} \lesssim I <
 I^{c_2}$. As shown in Fig.~\ref{FDynMRSS_Ma}c, the maximum period is in the
 form $2^j (2^M - 1)$ where $j$ is an integer greater than or equal to $2$.
\end{itemize}

Our findings of the dynamics of the minority choice time series can be
tabulated in Table~\ref{Tperiod}.

\begin{table}[ht]
 \begin{tabular}{|c|c|c|c|c|c|}
  \hline
   Period of & MG & \multicolumn{2}{|c|}{HMG$^\text{FSS}$} &
    \multicolumn{2}{|c|}{HMG$^\text{MRSS}$} \\
  \cline{3-4} \cline{5-6}
   dynamics & & $I < I^{c_1}$ & $I > I^{c_1}$ & $I \ll I^{c_1}$ &
    $I \gtrsim I^{c_1}$ \\
  \hline
   $\alpha \gg \alpha^\text{MG}_c$ & nil & nil & nil & nil & nil \\
  \hline
   $\alpha \approx \alpha^\text{MG}_c$ & $2^{M+1}$ & $2^{M+1}$ & nil &
    $2^{M+1}$ & intermittent\footnotemark[1] \\
  \hline
   $\alpha \ll \alpha^\text{MG}_c$ & $2^{M+1}$ & $2^{M+1}$ & nil & $2^{M+1}$ &
    nil\footnotemark[2]
    \footnotetext[1]{The maximum period of the orderly phase of this dynamics
     is $2^j (2^M - 1)$ where the value of $j$ can be found in the main text.}
    \footnotetext[2]{But the minority choice time series conditioned on an
     arbitrary but fixed history shows a very weak period two dynamics}
   \\
  \hline
 \end{tabular}
 \caption{Summary of the dynamics in the minority choice time series for MG and
  HMG for odd number of agents and $M\geq 2$.}
 \label{Tperiod}
\end{table}

\subsection{Conditions for the period $2^j\,\text{L.C.M.}(p_1,p_2,
 \ldots , p_M)$ dynamics in the orderly phase of the HMG$^\text{MRSS}$
 intermittency}
We find that in HMG$^\text{MRSS}$, some agents seldom change their strategies
while others do so frequently. We say that an agent is
{\bf oscillating} if he switches strategy within the previous $2^{M+1}$ turns.
Otherwise, the agent is said to be {\bf frozen}.
It turns out that the number of frozen agents and their performance provide
important information to allow us to understand the origin of the dynamics in
the minority choice time series of HMG$^\text{MRSS}$ for $I \gtrsim I^{c_1}$
and $\alpha \approx \alpha^\text{MG}_c$.

\begin{table}[ht]
\begin{tabular}{|c|c|c|}
 \hline
 & \multicolumn{2}{|c|}{\parbox{4cm}{mean winning probability}} \\
 \cline{2-3}
 $N$ & frozen agents & oscillating agents \\
 \hline
 4001 & 0.499 & 0.292 \\
 \hline
 8001 & 0.499 & 0.302 \\
 \hline
 16001 & 0.499 & 0.293 \\
 \hline
\end{tabular}
\caption{The winning probability of frozen and oscillating agents for $M=3$ and
 $I = 0.90$ averaged over 100 independent runs.}
\label{Table:B}
\end{table}

Table~\ref{Table:B} shows the average probabilities for a frozen (an
oscillating) agent to correctly predict the minority side in a typical
HMG$^\text{MRSS}$ with $I \gtrsim I^{c_1}$ and $\alpha \approx
\alpha^\text{MG}_c$. Clearly, a frozen (an oscillating) agent will have a
higher (lower) chance to
correctly predict the minority side (that is, the winning probability) in the
next turn. Since an oscillating agent must lose frequently in recent turns, our
finding means that badly performing agents are likely to perform badly in
future. More importantly, numerical simulations tell us that the presence of
the period $2^j \,\text{L.C.M.} (p_1,p_2, \ldots , p_M)$ dynamics in the
orderly phase of the intermittency
where $j \geq 1$ is an integer and $p_1, \ldots , p_M$ divide $(2^M - 1)$ is
almost always accompanied by the presence of only one oscillating agent in the
entire system. Besides, this oscillating agent uses only two distinct
strategies that are anti-correlated. (That is, these two strategies predict
different minority side for all given histories \cite{HJHJ01,HJHJ01a}.)
Moreover, the strategies used by the remaining frozen agents form a collection
of crowd-anticrowd pairs. Interestingly, our observed intermittency disappears
and becomes a single aperiodic phase if we replace the histories by random
variables or if the number of agents $N$ is even. Also, the sole oscillating
agent associated with each orderly phase of the time series may be different.

\section{The crowd-anticrowd explanation} \label{Explain}

Let us briefly review how the crowd-anticrowd theory explains the behavior of
the standard MG before adapting it to explain the behavior of HMG$^\text{FSS}$
and HMG$^\text{MRSS}$.

\subsection{The crowd-anticrowd theory for the standard MG}
According to the crowd-anticrowd theory, agent cooperation in the standard MG
is determined by the number of (effective) anti-correlated pairs of strategies
in current use. The smaller the difference between the number of agents
currently adopting a strategy $a$ and those currently adopting its
anti-correlated strategy $\bar{a}$, the better the crowd-anticrowd
cancellation leading to a better agent cooperation. Since the number of
available strategies is much less than the strategy space size for $\alpha
\gg \alpha^\text{MG}_c$, crowd-anticrowd cancellation cannot be effective in
this regime. And because standard MG agents do not have inertia, they switch
strategies immediately once the maximum virtual score difference is negative.
Due to the fact that the virtual score of a strategy in the standard MG is
independent of who owns or uses it, every standard MG agent has the same view
on the performance of a given strategy. So, when $\alpha \ll
\alpha^\text{MG}_c$, standard MG agents tend to adopt and drop similar
strategies all the time. This over-reaction leads to a herd effect and is the
origin of the maladaptation in the standard MG in this regime.
Thus, effective agent cooperation is possible only for
$\alpha \approx \alpha^\text{MG}_c$ in the standard MG. A remarkable
feature of the standard MG is that effective agent cooperation is indeed
possible in this regime in spite of the fact that agents act independently
by utilizing common global coarse-grained information only
\cite{HJHJ01,HJHJ01a}.

\subsection{Towards the crowd-anticrowd theory for HMG}
\label{Subsec:drivingforces}

Unlike the standard MG, HMG agents use Eq.~(\ref{EDOmega}) to decide whether
to keep their currently using strategies or not. In general, they are
initially assigned different strategies and begin to adopt their currently
using strategies at different times. So, they have different virtual score
difference $\Delta\Omega_k$ and the number of turns since the adoption of the
current strategy $\tau_k$.
Together with the virtual score reset mechanism stated in rule~\ref{HMGrule4}
of HMG, the same strategy may be ranked differently amongst HMG agents.
Consequently, by picking strategies from the same strategy space, the effective
strategy diversity for HMG is in general bigger than that for the standard MG.
Furthermore, the higher the value of $I$ (and hence the higher the inertia),
the slower the average rate of strategy
switching. All these factors reduce HMG agents' over-reaction and the herd
effect making the system to better cooperate \cite{MC06,CM07}.
Just like the standard MG, the dynamics of HMG encourages agents to form
crowd-anticrowd pairs thereby increasing agent cooperation.
But unlike the standard MG, this ``crowd-anticrowd pair formation''
driving force in HMG is more gentle and is less likely to cause over-reaction
and maladaptation because of the presence of inertia.
Hence, the higher the inertia, the longer the equilibration time.
At this point, we have to emphasize that the presence of inertia need not imply
that the agents must cooperate because the ``crowd-anticrowd pair formation''
driving force may not be strong enough in a certain parameter regime. The
bottom line is that the presence of inertia never worsen the agent
cooperation. In other words, for fixed values of $M$, $N$ and $S$, the
variance of attendance per agent $\sigma^2 / N$ for the standard MG has to be
greater than or equal to that for the HMG$^\text{FSS}$ or HMG$^\text{MRSS}$.

Suppose the agents really cooperate.
Still there are two ways to prevent them from cooperating forever.
Consider a history $\vec{\mu}$ which has non-zero probability of occurrence.
Suppose further that the minority choice time series conditioned on this
history $\vec{\mu}$ is biased. (That is, it is more likely to find a particular
minority choice than the other in this conditioned time series.) In this case,
certain strategy $a$ will outperform its anti-correlated partner $\bar{a}$ in
the long run. More precisely, the rate of change of virtual score difference
between $a$ and $\bar{a}$ averaged over a sufficiently long number of turns is
positive. So, after a sufficiently long time, agents will begin to drop
strategy $\bar{a}$ and adopt strategy $a$, making strategy $\bar{a}$ much more
popular than strategy $a$. In particular, if the minority choice time series
exhibits periodic dynamics, then the timescale for a frozen agent to change to
an oscillating agent via this mechanism is directly proportional to the period
of the dynamics and inversely proportional to the number of biased histories in
the minority choice time series.

Even if the minority choice time series conditioned on every history is
un-biased, there is still a way for agents to stop cooperating.
Provided that the value of $\Delta\Omega_k$ follows an un-biased random walk,
after sufficiently long time, agent~$k$ can switch his strategy once a while
due to fluctuations in $\Delta\Omega_k$. These two mechanisms act like a
``crowd-anticrowd pair destruction'' driving force that decreases agent
cooperation. Surely, the former mechanism is more efficient.

In summary, it is the combined actions of the above two driving forces that
determine the agent cooperation and dynamics of HMG. In fact, the
``crowd-anticrowd pair formation'' driving force dominates the initial dynamics
of the HMG. And the ``crowd-anticrowd pair destruction'' driving force becomes
important after most of the agents have been paired up. This picture allows us
to understand the simulation results reported in Sec.~\ref{Sec:Result}.

\subsection{The crowd-anticrowd explanation for HMG$^\text{MRSS}$}
\label{Subsec:CCExplanation}

In the case of $\alpha\gg \alpha^\text{MG}_c$,
there are so few strategies at play that most of the crowd-anticrowd pairs is
made up of only one agent. As a result, the ``crowd-anticrowd pair formation''
driving force is never strong enough to ensure agent cooperation
irrespective of the value of $I$. Thus, crowd-anticrowd cancellation is
ineffective. Besides, agents in effect make random choices each turn so that
the value of $\sigma^2 / N$ approaches the coin-toss limit as
$\alpha\rightarrow\infty$.
Surely, the minority choice time series does not show any periodic dynamics
\cite{HJHJ01,HJHJ01a}.
Since the above arguments are also valid for the standard MG and
HMG$^\text{FSS}$, we understand why MG, HMG$^\text{FSS}$ and HMG$^\text{MRSS}$
all behave in the same way in this parameter regime.

In the case of $I \lesssim I^{c_1}$,
the inertia of agents is so low that agents switch strategies immediately
whenever the maximum virtual score difference is negative. In other words,
the response of standard MG and HMG agents are the same in this parameter
regime \cite{MC06}. Hence, the agent cooperation and dynamics of standard MG,
HMG$^\text{FSS}$ and HMG$^\text{MRSS}$ are about the same.

The remaining case to study is $I \gtrsim I^{c_1}$ and
$\alpha \lesssim \alpha^\text{MG}_c$.
Note that the behavior of HMG$^\text{FSS}$ and HMG$^\text{MRSS}$ in this
case differ markedly as reported in Sec.~\ref{Sec:Result}.
We begin our analysis by stating the
following claim whose proof can be found in Appendix~\ref{AppendixCooperation}.

\begin{claim}
 Let $N$ be an even number. Suppose each of the $N$ players are randomly and
 independently assigned $S$ socks; and a sock has $2^M$ possible colors and can
 either be left or right. (Hence, there are $2^{M+1}$ kinds of socks.) Suppose
 further that each kind of sock is selected with equal probability. Then,
 provided that $2^{M+1} < N S$, the probability that there exists a way to form
 $N/2$ pairs of socks by picking exactly one sock from each of the $N$ players
 is greater than or equal to $1 - \beta / N$ for some positive $\beta$ which is
 independent of $N$. \label{claim}
\end{claim}

Recall that two strategies are said to be anti-correlated if they always
predict different minority side. And they are said to be uncorrelated if they
have equal chance to predict the same minority side provided that each history
occurs with equal probability. Thus, strategies in the MRSS consist of $2^M$
pairs of anti-correlated strategies and that strategies picked from two
distinct pairs are uncorrelated \cite{HJHJ01,HJHJ01a}.

By identifying the $2^{M+1}$ different kinds of socks with the $2^{M+1}$
different strategies in the MRSS, Claim~\ref{claim} implies that for a
sufficiently large $N$ and for $\alpha < 1$, the probability of forming
$\left\lfloor N/2 \right\rfloor$ pairs of anti-correlated strategies by picking
one strategy from each agent's strategy pool is high.
Surely, finding this solution requires communications amongst the agents.
If the agents could keep on using his particular choice of strategy throughout
the game, maximum agent cooperation would be attained and hence the
theoretical minimum value of $\sigma^2 / N$ (that is $\sigma^2 / N = 0$ if $N$
is even and $1/4N$ if $N$ is odd) would be resulted.
In contrast, since the strategy space size of the FSS is exponentially larger
than that of the MRSS in the large $M$ limit, the condition $\alpha < 1$ is not
sufficient for agents to maximally cooperate in the above way should they pick
strategies from the FSS. The correct condition in this case should be $2^{2^M}
< N S$ (and hence, $\alpha \rightarrow 0^+$ in the thermodynamic limit).

Claim~\ref{claim} only assures the existence of an optimal way of agent
cooperation with a high probability. It remains to show that this maximum
agent cooperation can be achieved with a high chance for HMG$^\text{MRSS}$
under certain conditions.
Recall from step~\ref{HMGrule5} of
the rules of HMG that agent~$k$ uses the maximum virtual score difference
$\Delta\Omega_k$ amongst all the strategies initially assigned to him to
decide whether to change strategy or not. The following consequences of
step~\ref{HMGrule5} are important to understand the strong dependence of HMG on
the strategy space used:
\begin{enumerate}
\item \label{con1}
The higher the value of $I_k$, the more willingly for agent~$k$ to stick to his
currently using strategy.
\item \label{con2} For fixed values of $I_k$'s, agents with a strategy and its
anti-correlated partner in their pool of strategies have the strongest
tendency, on average, to switch their strategies.
\item \label{con3} Suppose agent~$k$ has just switched to a new strategy and
that this newly adopted strategy incorrectly predicts the minority side in its
first use, then $\tau_k = 1$ and $\Delta\Omega_k = -2$. As a result, this agent
will drop his newly adopted strategy in the next turn provided that $I_k <
I^{c_2}$.
\end{enumerate}

\subsubsection{The sub-case of $I \gtrsim I^{c_1}$ and
$\alpha \approx \alpha^\text{MG}_c$}
Shortly after the commencement of HMG$^\text{MRSS}$, the minority choice time
series should resemble an un-biased random sequence. So provided that $I$ is
sufficiently large, Subsec.~\ref{Subsec:drivingforces} tells us that the
``crowd-anticrowd pair formation'' driving force allows a large number of
agents to form crowd-anticrowd pairs. Most of these paired agents will be
frozen, and there are only a few oscillating agents in the system. By simple
probability consideration, we expect that most of the strategies hold by these
oscillating agents are un-correlated. As the dynamics of the system,
which is determined mostly by the dynamics of these oscillating agents, the
minority choice time series conditioned on an arbitrary history is likely to be
un-biased. Hence, fewer and fewer oscillating agents will present as they
gradually form crowd-anticrowd pair and become frozen.
From Claim~\ref{claim}, we believe that for a sufficiently large $I$, 
agents in HMG$^\text{MRSS}$ have a high chance to attain maximum
agent cooperation provided that $\alpha \approx \alpha^\text{MG}_c < 1$.
This is consistent with the findings in our
numerical simulations reported earlier in Sec.~\ref{Sec:Result} that the
highest chance of finding maximum agent cooperation is when $\alpha \approx
\alpha^\text{MG}_c$ and $I \gtrsim I^{c_1}$. And this maximum agent
cooperation is accompanied by the existence of at most one oscillating agent in
the system who switches between a pair of anti-correlated strategies.
This finding agrees with the discussion following Claim~\ref{claim} that
the agent holding a pair of anti-correlated strategies switches his strategy
most readily.
To conclude, when $N$ is odd, the effective number of strategies at play for
HMG$^\text{MRSS}$ in this regime is reduced to one in most of the
time. Thus, the frozen agents have an average of 1/2 chance to correctly
predict the minority side in the next turn while the only oscillating
agent has no chance to do so. Consequently, the frozen agents are unlikely to
switch their currently using strategies while the only oscillating agent is
prone to strategy switching.

We now discuss the dynamics of the minority choice time series when $N$ is odd.
From the above discussions, it is clear that the system is in an orderly
(chaotic) phase whenever it has one (more than one) oscillating agent. In
addition, this oscillating agent
is most likely to be switching between a pair of anti-correlated strategies.
This oscillating agent always predicts the minority side incorrectly
throughout the corresponding orderly periodic dynamics phase. Thus, from the
discussions immediately after Claim~\ref{claim}, we conclude that this
oscillating agent must drop his currently using strategy and switch to its
anti-correlated counterpart in each turn provided that $I^{c_1} \lesssim I <
I^{c_2}$. By identifying a strategy with a linear function of the historical
minority choices $\mu_i$'s over the finite field $GF(2)$ in the form of
Eq.~(\ref{E:MRSS_Def}), we know that the difference between the linear
functions associated with two anti-correlated strategies is equal to $1$.
So, whenever $\alpha \approx \alpha^\text{MG}_c$ and $I^{c_1} \lesssim I <
I^{c_2}$, the minority side in the $n$th turn $\mu_n$
throughout this orderly periodic dynamics phase obeys
\begin{equation}\label{eq:RSS}
 \mu_n=\sum_{i=1}^M \eta_i \mu_{n-i} + \eta_0 + n ~,
\end{equation}
where $\eta_i, \mu_i \in GF(2)$ and $n$ denotes $1+1+ \cdots +1
\text{~($n$ terms)}$.

We show in Appendix~\ref{App:B} the following theorem.
\begin{thrm}
 The sequence $\{ \mu_n \}$ generated by the recursion relation in
 Eq.~(\ref{eq:RSS}) is periodic. Its period is in the form $2^j \text{L.C.M.}
 (p_1,p_2,\ldots , p_M)$ where $j\geq 1$, and $p_1, \ldots , p_M$ are positive
 integers dividing $(2^M - 1)$. Moreover, the longest possible period for this
 sequence is $2(2^M - 1)$ if $M\geq 2$ and $4$ if $M = 1$. Suppose the sequence
 is of maximum possible period and $M\geq 2$. Denote the history $(
 \mu_{n-M+1}, \mu_{n-M+2}, \ldots , \mu_{n-1}, \mu_n )$ by $\vec{\mu}(n)$.
 Then, the two histories of alternating $0$'s and $1$'s, namely,
 $(0,1,0,1,\ldots)$ and $(1,0,1,0,\ldots)$ appear in the sequence of history $H
 = \{ \vec{\mu} (n) \}_{n=1}^{2(2^M - 1)}$ once; while all the other
 $(2^M - 2)$ possible histories all appear in the history sequence $H$ twice.
 \label{Thrm}
\end{thrm}

This theorem allows us to explain the period of the orderly phase of the
time series in the case of $I^{c_1} \lesssim I < I^{c_2}$ and $\alpha \approx
\alpha^\text{MG}_c$. In particular, the proof tells us that the longest
periodic dynamics in the orderly phase for $M\geq 2$ is of period
$2(2^M - 1)$.
Nevertheless, periods satisfying Theorem~\ref{Thrm} do not show up in the
orderly phases equally frequently. If the number of turns between two
consecutive occurrence of a history in an orderly phase is odd, then the
minority time series conditioned on this history shows period two dynamics
because the sole oscillating agent makes alternating prediction of the minority
side each time when he is given the same history. In other words, the minority
choice time series conditioned on this history is un-biased. In contrast, if
the number of turns between two consecutive occurrence of a history in an
orderly phase is even, the minority time series conditioned on this history is
biased for it exhibits period one dynamics. Thus, the ``crowd-anticrowd
destruction'' driving force discussed in Subsec.~\ref{Subsec:drivingforces}
will break the maximum agent cooperation in a time
proportional to the period of this orderly phase and inversely proportional
to the number of conditional minority choice time series that exhibits period
one dynamics.

Theorem~\ref{Thrm} tells us that for $M\geq 2$, the longest possible period
of the orderly phase is $2(2^M - 1)$. This dynamics is present when the
(degree $M$) characteristic polynomial of the recursion
relation~(\ref{eq:RSS}), which is associated with the pair of anti-correlated
strategies used by the sole oscillating agent, is primitive. In a
single period,
the two histories that consists of alternating $0$'s and $1$'s appear once
while all other histories appear twice. Therefore, the number of turns between
two consecutive appearance of each of the two histories consisting of
alternating $0$'s and $1$'s equal the even number $2(2^M - 1)$. It is easy to
show that the number of turns between two consecutive appearance of all other
histories must be odd. (One way to do so is that if $\vec{\mu} (k) = \vec{\mu}
(k')$ with $(k - k')$ being a positive even number, the homogeneous parts of
the solutions of Eq.~(\ref{eq:RSS}) for $n = k - i$ and $n = k' - i$ are
equal whenever $i =
0,1,\ldots , M-1$. And this is possible only when $(2^M - 1) \mid (k - k')$.
As $(k - k')$ is even, so $k - k' \geq 2(2^M - 1)$. Since all other
histories occurs twice in a single period, the number of turns between two
consecutive occurrence for them must be odd.) In conclusion, the $2(2^M - 1)$
period dynamics can be found in the minority choice time series although it
cannot be ever-lasting. Since out of the $2^M$ possible minority choice time
series conditional on a particular history, only two of them show period
one rather than period two dynamics, the period $2(2^M - 1)$ dynamics in the
minority choice time series is quite stable in the sense that it
lasts for a longer time. This also explains why the stability of this period
increases with $M$.

As $M$ increases while keeping $S$ and $\alpha$ fixed, the
probability of having an agent in the system that holds a pair of
anti-correlated strategy is about $1 - [1 - S(S+1) / 2^{M+2}]^N \rightarrow 1 -
e^{-(S+1)/2\alpha}$. As $\alpha \approx \alpha^\text{MG}_c \approx 0.3$ in the
intermittent phase, it is not surprising that there is a high chance for the
sole oscillating agent in this phase to switch between two anti-correlated
strategies. Nonetheless, from the proof of Theorem~\ref{Thrm}, the
corresponding probability of having an agent in the system that holds a pair of
anti-correlated strategy that causes the period $2(2^M - 1)$ dynamics is about
$1 - [ 1 - S(S+1)\varphi (2^M - 1) / 2^{2(M+1)} M ]^N$, where $\varphi$ denotes
the Euler-Totient function. By prime number theorem, this probability is equal
to at least $1 - [ 1 - S(S+1)(2^M - 1)/2^{2(M+1)} M^{c+1} ]^N \rightarrow 1 -
e^{-(S+1)/2\alpha M^{c+1}}$ as $M\rightarrow\infty$, where $c$ is a positive
constant. This explains why as $M$ increases, there is general trend that the
period $2(2^M - 1)$ dynamics orderly phase occurs less frequently.

From Theorem~\ref{Thrm}, the second longest period in the orderly phase
is of period $4(2^{M-1}-1)$ for $M\geq 2$. This dynamics is associated with a
characteristic polynomial in the form $(\lambda-1)p(x)$ where $p(x)$ is a
primitive polynomial of degree $(M-1)$. Again, using the idea in the proof of
Theorem~\ref{Thrm}, it is easy to check that the period $4(2^{M-1}-1)$ orderly
phase is quite stable (although it is not as stable as the period $2(2^M - 1)$
dynamics) as all but four conditional minority choice time series exhibit
period one rather than period two dynamics.

For those orderly phases with shorter periods, the number of distinct histories
present is small so that the effective diversity of the strategies at play are
greatly reduced. Combined with the presence of, in general, a larger
proportion of period one conditional minority choice time series in this
dynamics, these shorter period dynamics are much less stable in the sense that
they can last for a much shorter time.

The same analysis can be applied to the case when $I\geq I^{c_2}$. In this
case, the only oscillating agent in the system switches his strategy once every
few turns. Applying the same analysis as in the proof of Theorem~\ref{Thrm}, it
can be shown that the maximum period of the orderly phase is $2^j (2^M - 1)$
where $j$ is the number of turns between the adoption and termination of a
strategy for the sole oscillating agent. In this situation, the strength of a
lot of period two dynamics in the conditional minority choice time series are
weakened making the dynamics in the orderly phase less pronounced.

Note that one of the factors making the above intermittent behavior possible is
that the number of agents $N$ is odd. In case $N$ is even,
there are equal number of agents choosing each side. Hence, all agents are
frozen and have 1/2 chance of correctly predicting the minority side.
And the minority choice time series does not show any periodicity. This is
exactly what we observe in our simulation.

Let us discuss more about the agent cooperation. Even for $\alpha \approx
\alpha^\text{MG}_c$ and $I$ sufficiently large, the average value of $\sigma^2
/ N$ is still a little bit higher than the theoretical minimum due to three
reasons. First, a few initial quenched disorders does not allow maximum agent
cooperation. Second, the system may be trapped in a non-maximally cooperative
state even though the initial quenched disorder allows maximum agent
cooperation. Third, as we have pointed out in
Subsec.~\ref{Subsec:drivingforces}, the aperiodic phase of the intermittency is
associated with the sub-maximal agent cooperation making the value of
$\sigma^2 / N$ averaged over initial quenched disorder greater than the
minimum value. So, it is not surprising to find that the values of $\sigma^2 /
N$ for HMG$^\text{MRSS}$ and HMG$^\text{FSS}$ are about the same for $\alpha
\approx \alpha^\text{MG}_c$ and $I \gtrsim I^{c_1}$.

\subsubsection{The sub-case of $I \gtrsim I^{c_1}$ and
$\alpha \ll \alpha^\text{MG}_c$}
For fixed $M$ and $S$, the number of agents $N$ increases as $\alpha$
decreases. From the discussions in Subsec.~\ref{Subsec:drivingforces}, we know
for a sufficiently large $N$, the timescale for all except at most one of the
agents to form crowd-anticrowd pairs is longer than the timescale for a pair of
crowd-anticrowd agents to break up. Therefore, as
$\alpha \ll \alpha^\text{MG}_c$, it is highly unlikely for the system to attain
maximum agent cooperation. Besides, by further increasing $\alpha$,
overcrowding of strategies is severe as more and more agents are using the
same strategy with similar virtual scores and number of turns since its
adoption. As a result, maladaptation and herd effect begin to appear.
These are the reasons why for $I \gtrsim I^{c_1}$, the value of $\sigma^2 / N$
starts to increase as $\alpha$ falls below about 0.1.
This is also the reason why the minority choice time series conditioned on an
arbitrary but fixed history exhibits a very weak period two dynamics.
However this period two dynamics is much weaker than the one observed in the
standard MG because the presence of inertia makes the agents to response less
readily. As the effective strategy diversity of HMG$^\text{FSS}$ is greater
than that of HMG$^\text{MRSS}$ which is in turn greater than that of the
standard MG, the value of $\sigma^2 / N$ for HMG$^\text{FSS}$ is smaller than
that of HMG$^\text{MRSS}$ which is in turn smaller than that of the standard
MG provided that $I \gtrsim I^{c_1}$.

Discussions in the previous two sub-cases predict that by decreasing
$\alpha$ below $\approx 2^{2^M} / NS$, the value of $\sigma^2 / N$ for
HMG$^\text{FSS}$ will start to increase due to overcrowding of strategies.
Unfortunately, we are not able to check the correctness of our prediction
because the memory and run time requirements are too high.

\section{Discussions}
\label{Final}
In summary, we have performed extensive numerical simulations to study the
behavior of HMG$^\text{MRSS}$. We
found that HMG agents cooperate better provided that their strategies are
picked from the FSS instead of from the MRSS. This is because the effective
diversity of strategies is higher in the former case.
Based on the crowd-anticrowd theory \cite{HJHJ01,HJHJ01a},
we understood the origin of agent cooperation for HMG$^\text{MRSS}$
in various parameter ranges
by studying the interplay between the so-called ``crowd-anticrowd pair
formation'' driving force and the so-called ``crowd-anticrowd pair
destruction'' driving force. And we found that the difference in
the cooperative and dynamical behavior between HMG$^\text{FSS}$ and
HMG$^\text{MRSS}$ is mainly due to the structure of the strategy space used.
In particular, we were able to explain the novel
intermittent behavior of the system and the orderly phase dynamics of the
minority choice time series when $\alpha \approx \alpha^\text{MG}_c$,
$I \gtrsim I^{c_1}$, $N$ is odd and $\sigma^2 / N \approx 1/4N$.
Essentially, this novel orderly dynamics in the intermittent phase is caused by
the fact that the game is effectively reduced to a similar game played by only
one agent most of the time and is accompanied by the maximum possible agent
cooperation. And this reduction is possible in case
the strategy pool is MRSS rather than FSS.

On one hand, HMG$^\text{MRSS}$ appears to be special in the sense that it is
the only variant of MG we have examined whose behavior depends
sensitively on whether the FSS or the MRSS is used.
On the other hand, the assumption that the behavior of
standard MG when played in FSS or in MRSS is about the same is only a working
assumption supported by numerical simulation results and heuristic
arguments \cite{HJHJ01,HJHJ01a}.
Although this assumption greatly simplifies the space complexity of numerical
simulations and sometimes even enable us to obtain a few
semi-analytical results, one should bear in mind that this is only an
assumption after all.
In this regard, it is instructive to study the
conditions under which one can replace FSS by MRSS without
significantly affecting the dynamics and cooperative behavior of a system.

Lastly, we remark that the standard way of using the number
$2^{M+1}$ to measure the diversity of strategies is no longer appropriate
for HMG. We believe that by suitably defining the diversity of strategies
(and hence also the expression for $\alpha$), the $\sigma^2 / N$ against
$\alpha$ curves for HMG played using different choices of strategy spaces can
be made to coincide, at least roughly.

\begin{acknowledgments}
We thank the Computer Center of HKU for their helpful support in providing the
use of the HPCPOWER system for simulations reported in this paper.
\end{acknowledgments}

\appendix
\section{Proof of claim~\ref{claim}}
\label{AppendixCooperation}

Let $\Gamma, \Gamma'$ denote two different partitions of the $N$ players into
$N/2$ pairs. Let $\text{Pr}_\Gamma$ denotes the probability that it is possible
for each of the $N/2$ pairs of agents in the partition $\Gamma$ to pick a sock
of the same color but different side. Clearly, $\text{Pr}_\Gamma$'s are
exchangeable. That is,
$\text{Pr}_\Gamma = \text{Pr}_{\Gamma'}$. In fact, for $S\ll 2^{M+1}$,
\begin{eqnarray}
 \text{Pr}_\Gamma & \approx & \left\{ \frac{\left( 2^{M+1} \right)
  !}{2^{S(M+1)} (2^{M+1} - S)!} \left[ 1 - \left( 1 - \frac{S}{2^{M+1}}
  \right)^S \right] \right\}^\frac{N}{2} \nonumber \\
 & \approx & \frac{S^N}{2^{M N S}} ~. \label{E:Pr_eq}
\end{eqnarray}

By de~Finetti theorem \cite{deFinetti} and its generalization by Diaconis and
Freedman \cite{deFinetti2}, as long as $0 < \text{Pr}_\Gamma < 1$ for any
finite $N$, the probability that there exists a way to form $N/2$
pairs of socks by picking one sock from each player is lower bounded by
$\text{Pr}(\infty) - \beta / N$ for some $\beta > 0$ independent of $N$, where
$\text{Pr}(\infty)$ denotes the same probability when $N\rightarrow\infty$.

So, to prove this claim, it suffices to show that $\text{Pr}(\infty) = 1$
whenever $2^{M+1} < N S$ and $S\geq 2$. Under these two conditions, it is
obvious that $\text{Pr}_\Gamma \in (0,1)$ for any finite $N$. So, de~Finetti
theorem implies that $\text{Pr}_\Gamma$'s are conditionally independent given
the tail $\sigma$-field \cite{deFinetti}. Thus,
\begin{equation}
 \text{Pr}(\infty) = \lim_{N\rightarrow\infty} 1 - \prod_\Gamma ( 1 -
 \text{Pr}_\Gamma ) = 1 ~.
\end{equation}
Hence, claim~\ref{claim} is proved.
\hfill$\Box$
\par\medskip
Before leaving this appendix, we point out a common mistake people makes. The
probability that all paired players have matching socks for an arbitrary but
fixed partition $\Gamma$ is equal to $\text{P}(\Gamma) \approx \left[ 1 -
\left(1 - S/2^{M+1} \right)^S \right]^{N/2}$.  Thus, the expected number of
partitions satisfying this ``all paired players'' condition is
$\text{P}(\Gamma) \times N! / [ (N/2)! ]^2$. A threshold condition for the
existence of a partition that all paired players have matching socks can then
be deduced by saddle point approximation. The loophole in this argument is that
socks are assigned once and for all to the players so that the probabilities
$\text{P}(\Gamma)$ and $\text{P}(\Gamma')$ are not independent.

\section{Proof of Theorem~\ref{Thrm}}
\label{App:B}
This proof uses a number of finite field concepts and techniques.
Readers who are not familiar may consult Ref.~\cite{finitefield} before moving
on.
The characteristic equation of Eq.~(\ref{eq:RSS}) is 
\begin{equation}\label{eq:char}
\lambda^M - \sum_{i=1}^M \eta_i \lambda^{M-i} = 0 ~.
\end{equation}

If the characteristic equation has no degenerate root, the homogeneous part
of the solution of $\mu_n$ in Eq.~(\ref{eq:RSS}) is in the form $\sum_i g_i
w_i^n$ where $w_i$ are roots of Eq.~(\ref{eq:char}) over the extension field
$GF(2^M)$ for some $g_i\in GF(2^M)$. Clearly, the period of the sequence $\{
\sum_i g_i w_i^n \}_{n=0}^{+\infty}$ divides $\text{L.C.M.} (|w_1|,|w_2|,\ldots
,|w_M|)$, where $|w_i|$ denotes the order of $w_i$. And the equality holds if
$g_i$'s are all non-zero. (Remember also that $|w_i|$ divides $(2^M - 1)$ for
all $i$.)

What if the roots of Eq.~(\ref{eq:char}) are degenerate?
Suppose $w_1$ is a double root of Eq.~(\ref{eq:char}). Then, $w_1^n$ and
$n w_1^n$ are the two generators of the homogeneous part of the solution of
Eq.~(\ref{eq:RSS}). Besides, the period of the sequence $\{ g_1 w_1^n +
g_2 n w_1^n \}$ divides $2 |w_1|$ with the equality holds if $g_1, g_2 \neq 0$.
Similarly, let $2^j$ be the smallest integer greater than or equal to $k$.
Then, it is straight-forward to check that $2^j |w_1|$ is divisible by
the period of the homogeneous part of the solution of the recursion
relation~(\ref{eq:RSS}) that corresponds to a degree $k$ root $w_1$ of the
characteristic equation~(\ref{eq:char}). Furthermore, the period can attain the
value $2^j |w_1|$ provided that the coefficients $g_i$'s are non-zero.

Combining the two cases above, we conclude that the period of the homogeneous
solution of Eq.~(\ref{eq:RSS}) divides $2^j \text{L.C.M.} ( |w_1|, |w_2| ,
\ldots )$ where $w_i$'s are the distinct roots of its characteristic equation
and $j$ is a non-negative integer. The maximum possible period for this
homogeneous solution is $(2^M - 1)$; and this is attainable provided that the
coefficient in each term of the homogeneous solution is non-zero and the
characteristic polynomial of Eq.~(\ref{eq:RSS}) is any one of the
$\varphi(2^M-1)/M$ primitive polynomials of degree $M$ in the $GF(2)[\lambda]$,
where $\varphi$ is the Euler-Totient function. Note further that if the
minority choice time series were generated by the homogeneous solution of
Eq.~(\ref{eq:RSS}) alone, then this time series would not contain $M$
consecutive $0$'s should its period be $(2^M - 1)$. Besides, in one period, all
history strings but the one consists of all $0$'s would appear exactly once.

Now, we study the particular solution of Eq.~(\ref{eq:RSS}).
The first case to consider is when the characteristic polynomial has an odd
number of (non-zero) terms. (This includes the case when the polynomial is
irreducible and $M \geq 2$ for otherwise 1 is a root of Eq.~(\ref{eq:char}).)
In this case, a particular solution of Eq.~(\ref{eq:RSS}) is $\mu_n = \eta_0+n
\pmod 2$ if $|\{i:\eta_{n-2i}\neq0\}|$ is even or $\mu_n = \eta_0+n+1 \pmod 2$
if $|\{i:\eta_{n-2i} \neq0\}|$ is odd. In either case, the period of this
particular solution is $2$.

The remaining case is when the number of (non-zero) terms in
Eq.~(\ref{eq:char}) is even. In this case, $(\lambda-1)$ must be a factor of
the R.H.S. of Eq.~(\ref{eq:char}). We write the R.H.S. of Eq.~(\ref{eq:char})
as $(\lambda-1)^kp(\lambda)$ for some polynomial $p(\lambda)$ with $p(1)
\neq 0$. If $k = 1$, a particular solution is $\mu_n = \left\lfloor
(n + 1 - \eta_0)/2 \right\rfloor \pmod 2$ which is of period $4$. Similarly, it
is easy to check that there is a particular solution of Eq.~(\ref{eq:RSS})
whose period divides $2^j$ where $2^j$ is the smallest integer greater than
$(k+1)$.

By combining the studies of the periods of the homogeneous and particular
solutions of Eq.~(\ref{eq:RSS}), we conclude that the period of the general
solution of Eq.~(\ref{eq:RSS}) is in the form $2^j \text{L.C.M.} (|w_1|, |w_2|,
\ldots )$ where $w_i$'s are the distinct roots of Eq.~(\ref{eq:char}) and $2^j$
is the smallest integer greater than $(k+1)$.
Moreover, the maximum possible period for the solution of the recursion
relation~(\ref{eq:RSS}) is $2(2^M - 1)$ for $M\geq 2$ and $4$ for $M = 1$.
In the former case, such a maximum period is attained only if the
characteristic polynomial of Eq.~(\ref{eq:RSS}) is primitive. And in the latter
case, it is attained when the characteristic polynomial is $\lambda - 1$.

Finally, we consider the case when the period of the sequence $\{ \mu_n \}$
attains its maximum possible value of $2(2^M - 1)$. In this case, a
particular solution of Eq.~(\ref{eq:RSS}) corresponds to the histories of
alternating $0$'s and $1$'s. So, combined with our earlier analysis on the
frequency of histories for the homogeneous part of the solution of
Eq.~(\ref{eq:RSS}), we conclude that in each period, all but the two histories
of alternating $0$'s and $1$'s appear twice. Besides, the two histories of
alternating $0$'s and $1$'s appear once. This proves the theorem.
\hfill$\Box$

\bibliography{mg13.3}

\end{document}